\documentstyle[preprint,tighten,prd,aps,epsfig,floats]{revtex}
\def\eval#1{\left\langle#1\right\rangle}
\begin{document}
\preprint{hep-ph/9709228}
\title{Charm---a thermometer of the mixed phase\thanks{Talk delivered by B.S.
at {\em RHIC '97, The Brookhaven Theory Workshop on Relativistic Heavy Ions},
July 6--16, 1997.}}
\author{Benjamin Svetitsky and Asher Uziel}
\address{School of Physics and Astronomy, Raymond and Beverly Sackler
Faculty of Exact Sciences, Tel~Aviv~University, 69978~Tel~Aviv, Israel}
\date{3 September 1997}
\maketitle
\begin{abstract}
A charmed quark experiences drag and diffusion in the quark-gluon plasma,
as well as strong interaction with the plasma surface.
Our simulations indicate that charmed quarks created in heavy ion
collisions will be trapped in the mixed phase and will come to
equilibrium in it.
Their momentum distribution will thus reflect the temperature at
the confinement phase transition.
\end{abstract}
\section*{\ }
Consider charm created in a high-energy nuclear collision.
Some 99\% of the charmed quarks created in hadronic collisions are
{\em not} to be found in $c\bar c$ bound states, but rather in the open charm 
continuum.
Much like their bound counterparts, the unbound charmed quarks are created
early, move through the interaction region slowly, and react strongly
with their environment.
They should contain as much information about the collision region
as the $J/\psi$, although this information may be harder to extract.

Assuming invariance of the collision kinematics under longitudinal
boosts \cite{Bj}, a space-time picture shows that a charmed quark, created
near $t=z=0$ (the initial nucleus--nucleus collision), moves
according to $z=v_0t$ which is exactly a longitudinal streamline
of the fluid.
Thus there is not much information to be gained from the quark's
eventual longitudinal momentum;
it is to the transverse momentum that we turn in order to learn about
the quark's interaction with the surrounding matter.

The quark is created in the nascent plasma and, assuming it is not created
too near the edge, sees this plasma cool to the mixed phase and beyond,
to the hadron gas which dissociates soon after.
The mixed phase lasts a long time, typically ten times as long as the pure
plasma which precedes it.
Our calculation thus concentrates on the charmed quark's interaction
with the mixed phase \cite{bqsl}.
We track the quark's diffusion through the initial plasma, its hadronization
upon emerging into the hadron phase, its collisions (as a $D$ meson) with
pions, and its possible reabsorption by a plasma droplet.
For a given initial transverse momentum, we calculate the distribution
of the final $p_\perp$ of the $D$ meson.
(For details of the following see \cite{us}.)

Our simulation of the mixed phase is based on the cascade hydrodynamics
code of Bertsch, Gong, and McLerran \cite{Bertsch}.
At proper time $\tau=\tau_p$, the uniform plasma breaks up into droplets
of radius $r_0=1.0$~or~$1.5$~fm (a parameter of which we have little
knowledge).
Each droplet has a mass equal to its volume times the Stefan-Boltzmann
energy density at the transition temperature, $M=(\frac43\pi r_0^3)\times
(\gamma T^{*4})$, and its temperature is fixed at $T^*$.
Each droplet then radiates pions as a black body.
At each pion emission, the droplet shrinks (to conserve energy) and recoils
(to conserve momentum).
Similarly, a droplet absorbs any pion that comes close enough, with the
concomitant growth and recoil.
Pion--pion collisions are included in the simulation as well.
The dominant motion of the droplet--pion mixture is a rarefaction due
to the longitudinal expansion imposed by the initial conditions.
This leads to the eventual evaporation of the droplets.

A charmed quark is created inside the plasma at $\tau_0<\tau_p$ and
is allowed to diffuse according to a Langevin process \cite{bqs},
\begin{eqnarray*}
{\bf x}(t+\epsilon)&=&{\bf x}(t)+\frac{{\bf p}(t)}{E(t)}\epsilon\\
{\bf p}(t+\epsilon)&=&{\bf p}(t)-\gamma(T){\bf p}(t)\epsilon+\bbox{\eta}(t)
\sqrt{\epsilon}\ ,
\end{eqnarray*}
where $\bbox{\eta}(t)$ is a Gaussian noise variable.
The plasma cools as $T=T_i(\tau_i/\tau)^{1/3}$ until it reaches $T^*$,
whereupon it breaks into droplets.
(We make sure the quark is {\em inside} a droplet.)
The quark continues to diffuse.
If the quark hits the surface of a droplet, it stretches a flux tube
out into the vacuum.
This flux tube has a tension $\sigma=0.16~\text{GeV}^2$ and a fission
rate (per unit length) $d\Gamma/d\ell=0.5$~to~$2.5~\text{fm}^{-2}$.
(This is the range of values used in string-based event generator programs.)
If the flux tube breaks in time, then the $c$ quark finds itself to be
a $D$ meson outside the droplet; otherwise it is reflected back into the 
droplet.

If the quark indeed hadronizes, it is tracked as a $D$ meson in the
pion gas.
If this $D$ meson hits a droplet, it is absorbed:
The light quark gets stripped off and the $c$ quark proceeds back into the 
plasma.

Very often, a $c$ quark coming to the surface of its droplet has insufficient
energy to emerge as a meson.  (It needs about 350~MeV.)
In that case the quark is {\em trapped}.
Unless it gains energy through diffusion or through droplet
recoil, it will remain
trapped until the droplet evaporates away much later.

\begin{figure}
\centerline{\psfig{figure=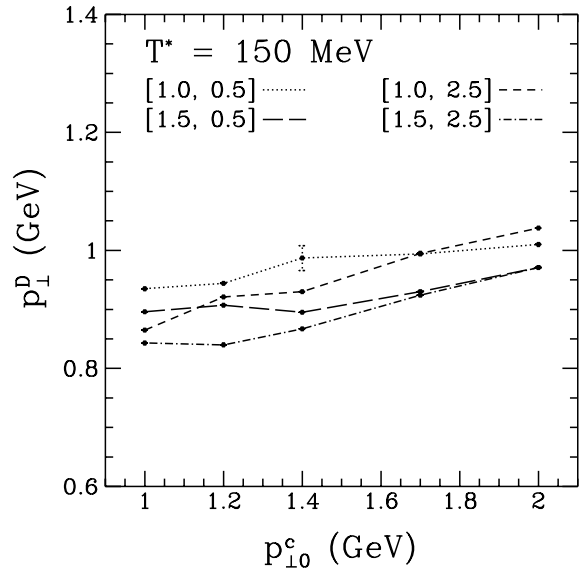,height=8cm}}
\caption{\label{fig1}
RMS transverse momentum of $D$ meson vs.~initial transverse momentum
of $c$ quark, for transition temperature $T^*=150$~MeV.
The four sets of points correspond to different values of $r_0$ and $d\Gamma/d\ell$
(in fm and fm$^{-2}$, respectively) as shown.
A typical statistical error bar is shown.}
\end{figure}

Now to results.
We show in Fig.~\ref{fig1} the mean $p_\perp$ of a $D$ meson as a function
of the initial $p_\perp$ of its parent $c$ quark.
Note the very weak dependence on the initial momentum.
This feature, together with the width of the final-momentum distribution,
indicates that the $D$ meson is {\em thermalized}.
The mean thermal $p_\perp$ of a $D$ meson at $T^*=150$~MeV is 820~MeV,
and we attribute the higher $\eval{p_\perp}$ shown in the figure to
the flow of the droplet fluid.

In interpreting our results we distinguish between two populations of
charmed quarks,
those that emerge from flux-tube fission and then escape the
system---{\em fragmentation mesons\/}---and
those that are trapped within their original droplets until the latter
evaporate---{\em breakup mesons\/}\@.
\begin{figure}\centerline{\psfig{figure=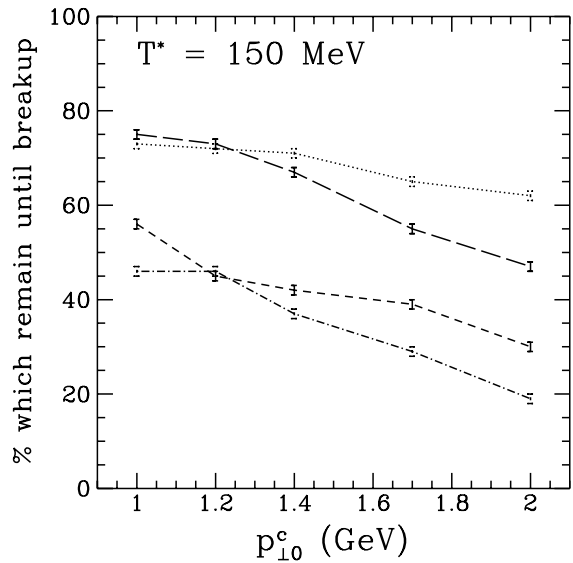,height=8cm}}
\caption{\label{fig2}The proportion of $c$ quarks which are trapped inside their
original droplets until their breakup.
The four sets of points are for different values of $r_0$ and $d\Gamma/d\ell$
as in Fig.~\protect\ref{fig1}.}
\end{figure}
As can be seen in Fig.~\ref{fig2}, between 40\% and 80\% of the quarks
are trapped until breakup.
\begin{figure}
\centerline{\psfig{figure=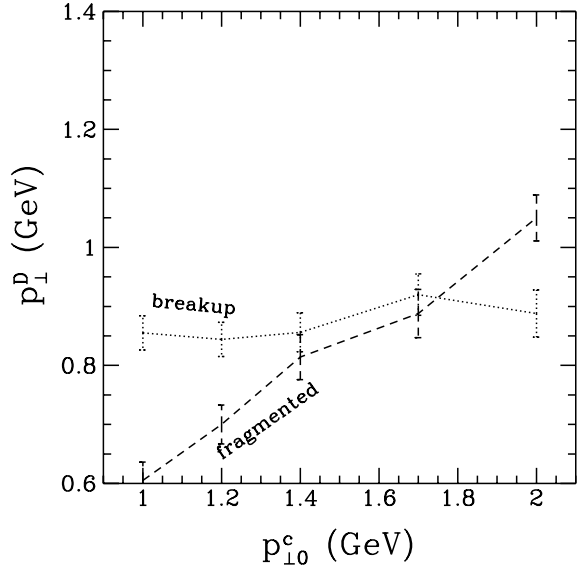,height=8cm}}
\caption{\label{fig3}
Comparison of two populations of $D$ mesons.
The RMS $p_\perp$ of $D$ mesons is plotted against the initial
$p_\perp$ of the $c$ quarks, for breakup mesons 
(dots) and for fragmentation mesons (dashes).
Here $T^*=150$~MeV, $r_0=1$~fm, $d\Gamma/d\ell=2.5$~fm$^{-2}$.}
\end{figure}
Fig.~\ref{fig3} shows that the breakup mesons are well thermalized while
the fragmentation mesons carry some memory of their quarks' initial
momentum.

\begin{figure}
\centerline{\psfig{figure=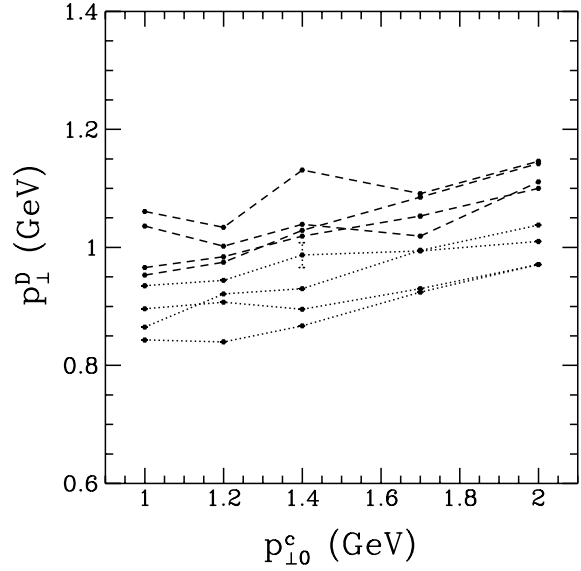,height=8cm}}
\caption{\label{fig4}
Comparison of the mean $D$ momentum for $T^*=150$~MeV (dots, same 
data as in
Fig.~\protect\ref{fig1}) with that for $T^*=200$~MeV (dashes).} 
\end{figure}

Finally, we compare results for two different transition temperatures.
Fig.~\ref{fig4} shows a comparison of the predictions for $T^*$=150~MeV with those
for $T^*$=200~MeV.
The $D$ transverse momentum shows a sensitivity to a temperature difference
of 50~MeV which rises above the effects of the
uncertainty in the two model parameters $r_0$ and $d\Gamma/d\ell$.

To summarize our qualitative conclusions:
\begin{enumerate}
\item The mean $p_\perp$ of a $D$ meson is almost independent of the initial
$p_\perp$ of the parent $c$ quark.
\item Many $c$ quarks (40\% -- 80\%) are trapped until the droplets evaporate,
and come to equilibrium at the transition temperature.
\item Other $D$ mesons come close to equilibrium in the plasma and in the
pion gas, and in any case decouple shortly after the transition is complete.
\item Recapture by the droplets (which we haven't discussed)
is an important effect, affecting the
equilibration of 10\% -- 40\% of the particles.
\end{enumerate}
The distribution of $D$ momenta is roughly thermal, but the mean
momentum is above the thermal average because of transverse flow.
Presumably the latter may be gauged by the momentum distributions of the lighter
species.
Everything considered, the $p_\perp$ of the $D$ mesons does provide a rough
thermometer of the phase transition.

\medskip
B.S. thanks Brookhaven National Laboratory for making possible his
participation in the RHIC '97 workshop.

\end{document}